\documentclass[prl,twocolumn,preprintnumbers,superscriptaddress,groupedaddress]{revtex4}\usepackage{graphicx,amssymb,amsbsy,amsfonts,amssymb,amsmath}
\usepackage{hyperref}
\usepackage{multirow}
\usepackage{stackrel}
\usepackage{tikz}
\usetikzlibrary{calc}

\newcommand{\be}{\begin{equation}}
\newcommand{\ee}{\end{equation}}

\newcommand{\bea}{\begin{eqnarray}}
\newcommand{\eea}{\end{eqnarray}}

\def\spa#1.#2{\left\langle#1\,#2\right\rangle}
\def\spb#1.#2{\left[#1\,#2\right]}
\def\spash#1.#2{\spa{\smash{#1}}.{\smash{#2}}}
\def\spbsh#1.#2{\spb{\smash{#1}}.{\smash{#2}}}
\def\sand#1.#2.#3{%
\left\langle\smash{#1}{\vphantom1}^{-}\right|{#2}%
\left|\smash{#3}{\vphantom1}^{-}\right\rangle}
\def\sandpp#1.#2.#3{%
\left\langle\smash{#1}{\vphantom1}^{+}\right|{#2}%
\left|\smash{#3}{\vphantom1}^{+}\right\rangle}
\def\sandpm#1.#2.#3{%
\left\langle\smash{#1}{\vphantom1}^{+}\right|{#2}%
\left|\smash{#3}{\vphantom1}^{-}\right\rangle}
\def\sandmp#1.#2.#3{%
\left\langle\smash{#1}{\vphantom1}^{-}\right|{#2}%
\left|\smash{#3}{\vphantom1}^{+}\right\rangle}

\def\NC{{N_c}}

\def\Mathematica{{\sc Mathematica}}

\newbox\charbox
\newbox\slabox
\def\s#1{{      
        \setbox\charbox=\hbox{$#1$}
        \setbox\slabox=\hbox{$/$}
        \dimen\charbox=\ht\slabox
        \advance\dimen\charbox by -\dp\slabox
        \advance\dimen\charbox by -\ht\charbox
        \advance\dimen\charbox by \dp\charbox
        \divide\dimen\charbox by 2
        \raise-\dimen\charbox\hbox to \wd\charbox{\hss/\hss}
        \llap{$#1$} }}

\begin{document}

\title{
\boldmath
    Analytic Form of the Planar Two-Loop Five-Gluon Scattering Amplitudes in QCD
\unboldmath
}
\preprint{CP3-18-72, FR-PHENO-2018-014, IPhT-18/149}

\author{S.~Abreu}
\affiliation{Center for Cosmology, Particle Physics and
Phenomenology (CP3), Universit\'{e} Catholique de Louvain, 1348
Louvain-La-Neuve, Belgium}
\author{J.~Dormans}
\affiliation{Physikalisches Institut,
Albert-Ludwigs-Universit\"at Freiburg,
D--79104 Freiburg, Germany}
\author{F.~Febres Cordero}
\affiliation{Physikalisches Institut,
Albert-Ludwigs-Universit\"at Freiburg,
D--79104 Freiburg, Germany}
\affiliation{Physics Department, Florida State University
Tallahassee, FL 32306, U.S.A.}
\author{H.~Ita}
\affiliation{Physikalisches Institut,
Albert-Ludwigs-Universit\"at Freiburg,
D--79104 Freiburg, Germany}
\author{B.~Page}
\affiliation{Institut de Physique Th\'eorique, CEA, CNRS, Universit\'e Paris-Saclay, F-91191 Gif-sur-Yvette cedex, France}

\begin{abstract}
We present the analytic form of the two-loop five-gluon 
scattering amplitudes in QCD for a complete set of
independent helicity configurations of external gluons.
These include the first analytic results for five-point
two-loop amplitudes relevant for the computation of
next-to-next-to-leading-order QCD corrections at hadron 
colliders.
The results were obtained by reconstructing analytic expressions
from numerical evaluations. The complexity of the
computation is reduced by exploiting physical and analytical
properties of the amplitudes, employing a minimal basis 
of so-called pentagon functions that have recently 
been classified.
\end{abstract}

\maketitle

Scattering amplitudes in QCD are beautiful mathematical objects
that condense the physics of particle collisions. They play a
central role in providing theoretical predictions for
experimental observables, such as those measured
in particle collisions at the Large Hadron Collider at CERN, 
making higher-order quantum corrections highly desirable. The
higher the corrections one considers, the more intricate
the amplitudes become, needing to accommodate richer
unitarity and factorization properties.  Nevertheless, the 
analytic properties are so constraining that final results 
appear very compact in spite of the large computational effort required to obtain them.
In this work, we exploit these
properties to directly access the compact analytic results 
from numerical evaluations.

Analytic expressions for two-loop four-gluon amplitudes have
been known for more than a decade
\cite{Glover:2001af,Bern:2002tk}.
At higher multiplicity, analytic results have been obtained
for all-plus helicity amplitudes up to
seven points~\cite{Badger:2013gxa,Badger:2015lda,
Gehrmann:2015bfy,Badger:2016ozq,Dunbar:2016aux,Dunbar:2016gjb} 
and up to five points for the single-minus 
helicity configuration \cite{Badger:2018enw},
reaching a complexity where traditional methods start to break
down. Conventional approaches to analytic calculations have
seen considerable progress~\cite{Boels:2018nrr,Chawdhry:2018awn}
but involve large intermediate expressions, 
the size of which 
is highly sensitive to details of the approach and
obscures the simplicity of the result.
To sidestep this issue and make the most of recent progress on
five-point two-loop integral calculations 
\cite{Papadopoulos:2015jft,Gehrmann:2018yef,Abreu:2018rcw,Chicherin:2018mue}, two groups have
used numerical approaches to compute five-parton two-loop QCD
scattering amplitudes \cite{Badger:2017jhb,
Abreu:2017hqn,Badger:2018gip,Abreu:2018jgq}. Numerical
approaches are more resilient to the swelling of
intermediate expressions, and can be improved with
finite-field calculations to provide exact results
\cite{vonManteuffel:2014ixa}.
While numerical results are sufficient for Monte Carlo
integration, analytic results increase the evaluation efficiency
and facilitate the study of the mathematical structure of 
amplitudes.
It has been pointed out that the analytic form of
rational functions in scattering amplitudes can be
reconstructed from numerical samples
\cite{vonManteuffel:2014ixa,Peraro:2016wsq}
as already demonstrated for the two-loop four-point amplitudes
\cite{Abreu:2017xsl} and recently for the single-minus
five-gluon amplitude \cite{Badger:2018enw}.
Combined with the better understanding
of the analytic properties of the functions appearing in
loop amplitudes that was developed over the last few years, the
reconstruction of analytic expressions from numerical samples
circumvents the prohibitively 
large intermediate analytic steps
and directly targets the simpler final results.

In this letter, we use these techniques to compute a
complete set of analytic two-loop five-gluon
amplitudes in the limit of a large number of colors $N_c$.
We apply the extension of the numerical variant
\cite{Ossola:2006us,Ellis:2007br,Giele:2008ve,Berger:2008sj}
of the unitarity method~\cite{Bern:1994zx,Bern:1994cg,
Britto:2004nc} to multi-loop computations
\cite{Ita:2015tya,Abreu:2017xsl,Abreu:2017idw} to obtain 
numerical samples on finite fields based on 
\cite{Abreu:2017hqn,Abreu:2018jgq}.
From these samples we obtain analytic 
expressions for the coefficients in a decomposition of
the amplitude in terms of so-called pentagon functions~\cite{Gehrmann:2018yef}.
The efficiency of the reconstruction is increased by finding
convenient bases in the space of pentagon functions and by an a
priori knowledge of the coefficients of the denominators.

We present for the first time analytic results
for the two-loop five-gluon amplitudes
known as MHV amplitudes,
which are an important ingredient in the next-to-next-to-leading-order 
(NNLO)
QCD corrections for three-jet production at hadron colliders.
The results will serve as a testing ground for
exploring the structure of two-loop scattering amplitudes in
quantum field theory.

\section{Scattering Amplitudes}

The main results of this letter are analytic expressions for
two-loop five-gluon scattering amplitudes, in
the leading-color approximation of pure Yang-Mills theory.
We write the perturbative expansion
of a bare five-gluon helicity amplitude as
\begin{align}\begin{split}\label{eq:bareAmp}
		&\mathcal{A}(\{p_i,h_i\}_{i=1,\ldots,5})\big\vert_{\textrm{leading color}} = \\
		&
\sum_{\sigma\in S_5/Z_5} {\rm Tr}\left(
    T^{a_{\sigma(1)}} T^{a_{\sigma(2)}}
    T^{a_{\sigma(3)}} T^{a_{\sigma(4)}} T^{a_{\sigma(5)}} \right)\\
		& g^{3}_0
    \left(\mathcal{A}^{(0)}
  +
		\lambda
		\mathcal{A}^{(1)} +
		\lambda^2\mathcal{A}^{(2)}
  +\mathcal{O}(\lambda^3)
  \right)\,,
\end{split}\end{align}
where $\lambda={\NC g_0^2}/{(4\pi)^2}$ and $g_0$ is the
bare QCD coupling.
The set $S_5/Z_5$ denotes all non-cyclic permutations of five
indices.
The planar amplitudes $\mathcal{A}^{(k)}$ are functions of the
momenta $p_{\sigma(i)}$ and helicities $h_{\sigma(i)}$.
In the leading-color approximation
there is a single color structure and it is
sufficient to specify a helicity assignment for the ordered set
of legs,
which we specify by subscripts 
$$\mathcal{A}^{(k)}_{h_1h_2h_3h_4h_5}$$ when
required, assuming an ordered set of momentum assignments 
$\{p_1,p_2,p_3,p_4,p_5\}$.
In this section we describe our approach to the calculation of 
the $\mathcal{A}^{(k)}$ in 
dimensional regularization with $D=4-2\epsilon$.
Although we focus on the
leading-color contributions, the approach we use is fully 
generic and applicable beyond this approximation.

Our calculation of the amplitudes $\mathcal{A}^{(k)}$
is performed in the framework of two-loop numerical
unitarity 
\cite{Ita:2015tya,Abreu:2017xsl,Abreu:2017hqn,Abreu:2017idw}.
The starting point is the general
parametrization of the integrand of the
amplitude~\cite{Ita:2015tya}
\begin{equation}\label{eq:AL}
    \mathcal{A}^{(2)}(\ell_l)=\sum_{\Gamma\in\Delta}
    \sum_{i\in M_\Gamma\cup S_\Gamma} c_{\Gamma,i}
    \frac{m_{\Gamma,i}(\ell_l)}{\prod_{j\in
    P_\Gamma}\rho_j},
\end{equation}
with $M_\Gamma$ a set of master integrands, $S_\Gamma$ a
set of surface terms, $P_\Gamma$ the set of propagators $\rho_j$
associated with propagator structure $\Gamma$, and $\ell_l$ the
set
of loop momenta. The coefficients $c_{\Gamma,i}$ are
determined by exploiting the factorization properties of
$\mathcal{A}(\ell_l)$ in loop-momenta limits 
$\ell_l \to \ell^\Gamma_l$
where propagators go
on-shell \footnote{At two-loops,
there are subleading poles in this limit which we handle
with the approach described in Ref.~\cite{Abreu:2017idw}.}:
\begin{equation}
\sum_{\rm states}\prod_{i\in T_\Gamma}
{\cal A}^{\rm tree}_i(\ell_l^\Gamma) =\!\!\!
\sum_{\substack{\Gamma' \ge \Gamma\,,\\ 
i\,\in\,M_{\Gamma'}\cup S_{\Gamma'}}} 
\frac{ c_{\Gamma',i}\,m_{\Gamma',i}(\ell_l^\Gamma)}{\prod_{j\in
(P_{\Gamma'}\setminus P_\Gamma) } \rho_j(\ell_l^\Gamma)}\,.
\label{eq:CE}
\end{equation}
The set $T_\Gamma$ labels all tree amplitudes corresponding to
the vertices in the diagram~$\Gamma$.
The sum runs over the $\Gamma'$ with more propagators.
The state sum runs over
the (scheme dependent) gluon-helicity states of each
internal line of $\Gamma$.
Solving the linear system in Eqn.~\eqref{eq:CE} directly yields
the coefficients of master integrals and no further integral
reduction is needed. In numerical unitarity, the linear system
\eqref{eq:CE} is constructed and solved numerically.
Using finite-field arithmetic removes any issues
related to loss of precision in numerical operations, at the
price of minor modifications to the standard numerical 
unitarity approach. We describe our algorithm in 
\cite{Abreu:2017hqn}.

\section{Numerical Reconstruction}
We aim to compute analytic expressions for five-gluon
amplitudes by reconstructing them from numerical samples. For a
suitably chosen parametrization of phase space, such as
momentum twistors \cite{Hodges:2009hk},
the coefficients $c_{\Gamma,i}$ of
Eqn.~\eqref{eq:AL} are rational functions. For concreteness,
we use the parametrization \cite{Peraro:2016wsq}
{\small 
\begin{align}\label{eq:twistorVars}
  &s_{12}=x_4,\,\,\, s_{23}=x_2 x_4,\,\,\,
  s_{34}=x_4\left(\frac{(1+x_1)x_2}{x_0}+x_1(x_3-1)\right),
  \nonumber\\
  &s_{45}=x_3x_4,\,\,\,
  s_{51}=x_1x_4(x_0-x_2+x_3)\,,\\
  &\textrm{tr}_5=i\,\varepsilon(p_1,p_2,p_3,p_4)
  \nonumber\\
  &=x_4^2\left(x_2(1+2x_1)+x_0x_1(x_3-1)
  -\frac{x_2(1+x_2)(x_2-x_3)}{x_0}
  \right),\nonumber
\end{align}
}
\!\!where $s_{ij}=(p_i+p_j)^2$ with the indices defined
cyclically.
The difficulty of analytic reconstruction is determined by the
complexity of the functions under study. More precisely,
this means that one should attempt to reconstruct rational
functions with low total degree of numerator and
denominator polynomials. With this aim
in mind, we first exploit a series of known
physical and analytical properties of the amplitudes, and
only apply the reconstruction algorithms to simpler objects 
that we cannot further constrain. In this section we summarize
our approach.

The divergence structure of scattering
amplitudes is governed by known universal functions
\cite{Catani:1998bh,Sterman:2002qn,Becher:2009cu,
Gardi:2009qi}.
It is thus sufficient to compute the so-called 
finite \emph{remainder} \cite{Catani:1998bh}, 
defined by subtracting contributions
that are related to tree and one-loop amplitudes
from a two-loop amplitude.
There is freedom in how the remainders are defined, so we now
give our definitions. For
helicity amplitudes which vanish at tree-level,
$\mathcal{A}^{(k)}_{\pm++++}$,
we use
\begin{equation}\label{eq:remainderHV}
  \mathcal{R}^{(2)}_{\pm++++}\!
	=\bar{\mathcal{A}}^{(2)}_{\pm++++}\!
  -\bar{\mathcal{A}}^{(1)}_{\pm++++}
  \sum_{i=1}^5
  \frac{(-s_{i,i+1})^{-\epsilon}}{\epsilon^2}
  +\mathcal{O}(\epsilon).
\end{equation}
The $\bar{\mathcal{A}}^{(k)}$ denote amplitudes
normalized to remove any ambiguity related to overall phases. In
the case of amplitudes that vanish at tree level, we
normalize to the leading order in $\epsilon$ of the (finite) 
one-loop amplitude.
For the MHV amplitudes, 
$\mathcal{A}^{(k)}_{-\mp\pm++}$,
which we normalize to the corresponding tree amplitude,
we define
\begin{align}\begin{split}\label{eq:remainderMHV}
  &\mathcal{R}^{(2)}_{-\mp\pm++}=
		\bar{\mathcal{A}}^{(2)}_{-\mp\pm++}
    -\left(\frac{5\,\tilde\beta_0}{2\epsilon}+{\bf I}^{
    (1)}\right)
  S_{\epsilon}\bar{\mathcal{A}}^{(1)}_{-\mp\pm++}
  \\
  &+    
 \left(
  \frac{15\,\tilde\beta_0^2}{8\epsilon^2}
  +\frac{3}{2\epsilon}
  \left(\tilde\beta_0{\bf I}^{(1)}-\tilde\beta_1
  \right)
  -{\bf I}^{(2)}
  \right)
  S_{\epsilon}^2
  +\mathcal{O}(\epsilon)\,,
\end{split}\end{align}
where $\tilde\beta_i$ are the coefficients of the QCD
$\beta$-function divided by $N_c$ and $S_\epsilon=(4\pi)^
{\epsilon}e^{-\epsilon\gamma_E}$. 
${\bf I}^{(1)}$ and ${\bf I}^{(2)}$ are the standard Catani
operators at leading color, 
and Eqn.~\eqref{eq:remainderMHV} is obtained by
writing Catani's prediction for the poles of a bare amplitude.
Precise expressions for the operators in our
conventions can be found in Appendix B of 
Ref.~\cite{Abreu:2018jgq}.
We note that in both Eqns.~\eqref{eq:remainderHV} and 
\eqref{eq:remainderMHV} we require one-loop
amplitudes expanded up to order $\epsilon^2$.

The amplitudes we are concerned with in this
letter can be written as linear combinations of special
functions, the so-called multiple polylogarithms (MPLs). The
coefficients in this linear combination are functions
of the external data and of the dimensional regulator
$\epsilon$, just as in the
decomposition of Eqn.~\eqref{eq:AL}. It is by now well
understood that there are relations between different MPLs, and
newly developed mathematical tools allow to find such relations
in an algorithmic way 
\cite{Goncharov:2010jf,Duhr:2011zq,Duhr:2012fh}.
It is then possible to construct a basis for the space of
special functions required for planar five-point massless
amplitudes up to two-loops, and this was done in 
\cite{Gehrmann:2018yef}
where a set of \emph{pentagon functions} was defined.
Amplitudes can then be written as
\begin{equation}\label{eq:ampAsPent}
  \bar{\mathcal{A}}^{(2)}=
  \sum_{i \in B}\sum_{k=-4}^0
  \epsilon^k\,\tilde c_{k,i}(\vec x)h_i(\vec x)\, + 
  \mathcal{O}(\epsilon),
\end{equation}
where $\vec x=\{x_0,x_1,x_2,x_3,x_4\}$ and the 
$\tilde c_{k,i}(\vec x)$
are rational functions of the twistor variables.
$B$ denotes the basis of pentagon functions $h_i$.
There is a notion of (transcendental) weight associated with
MPLs, and thus with pentagon functions, that allows to organize
the basis $B$. For instance, there is a single element of weight
0, the trivial function 1. At weight one, there are five
elements, $\ln(-s_{i,i+1})$. At weight two, there are products 
of weight-one functions, irreducible weight-two functions, and a new
constant, $\pi^2$. The same pattern holds at weight three and
four, the highest weight required for two-loop amplitudes at
order $\epsilon^0$.
To make the simplifications that one expects to find
in the remainders explicit, we write both one- and two-loop
amplitudes in terms of pentagon functions so that the remainders
themselves are written as combinations of pentagon functions:
\begin{equation}\label{eq:dec_as_masters}
\mathcal{R}^{(2)}=
\sum_{i \in B} r_i(\vec x)\,h_i(\vec x)\,.
\end{equation}

The coefficients $r_i(\vec x)$ have lower total degrees compared
with the $\tilde c_{k,i}(\vec x)$ of Eqn.~\eqref{eq:ampAsPent},
but to further increase the efficiency of the reconstruction
we find it useful to implement a series of improvements by
investigating the structure of the coefficients on
generic `univariate slices'. Such slices are defined by 
parametrizing the twistor variables $\vec x$ by a single 
variable $t$, $x_i=a_i+b_i\,t$, with the $a_i$ and $b_i$ chosen
randomly in the finite field (for high enough cardinality, this
ensures the $x_i$ do not satisfy simple relations leading
to artificial simplifications).
The coefficients $r_i(\vec x)$ are themselves univariate 
rational functions of $t$, $r_i(\vec x(t))=r_i(t)$.
Importantly, on such univariate slices the degrees of the 
numerator and denominator of the rational
functions $r_i(t)$ correspond to the total degrees of 
$r_i(\vec x)$  in the $x_j$.
We use these slices to probe the
complexity of the functions and find
simplifications.
Firstly, we classify the pole structures of the coefficient 
functions $r_i(\vec x)$. A similar classification has been 
exploited for the computation of one-loop QCD amplitudes in the 
past~\cite{Berger:2008sj}.
On physical grounds we expect the pole structure to be 
determined by the so-called alphabet of the pentagon functions.
The alphabet determines the points in phase space
where the pentagon functions (or their discontinuities) have
logarithmic singularities, and they provide a natural ansatz
for the poles of the coefficients.
The five-point planar alphabet can be written in terms of 26
letters $W_i$, see Ref.~\cite{Gehrmann:2018yef}, which we
rewrite in terms of twistor variables.
This gives a set $A$ of 26 independent letters 
$A=\{w_i(\vec x)\}$.
Indeed, we find that the denominators do factorize into
products of letters,
\begin{equation}\label{eq:ansatz}
r_i(\vec x) =  
\frac{n_i(\vec x)}{\prod_{j\in A} w_j(\vec x)^{q_{ij}}}\,.
\end{equation}
The integer exponents $q_{ij}$ are determined by matching 
this ansatz on univariate slices.
The computation of the $r_i(\vec x)$ is then reduced to 
the much simpler multi-variate polynomial reconstruction of the
numerators $n_i(\vec x)$.
Secondly, it is expected that cancellations
between different basis functions  in $B$ take
place in exceptional kinematic configurations, which implies 
relations between their coefficients.  
This motivates a search
for a different basis of pentagon functions with coefficients 
of lower total degree. To find this new (helicity-dependent) 
basis, we construct linear combinations of coefficients
\begin{equation}
  \sum_{i \in B} a_{i,k} r_i(\vec x)=
  \frac{N_k(\vec x,a_{i,k})}
  {\prod_{j\in A} w_j(\vec x)^{q_{kj}'}}\,,
\end{equation}
and solve for phase space independent $a_{i,k}$ such that the 
numerators $N_k(\vec x,a_{i,k})$ factorize a subset of the
$w_j\in A$.
This can be performed on univariate slices by
only accepting solutions which are invariant over multiple 
slices. The matrix $a_{i,k}$ allows to change to a new basis 
$B'$ in the space of special functions, in which remainders
can be decomposed as in Eqn.~\eqref{eq:dec_as_masters}, with
coefficients $r_i'(\vec x)$ whose numerators $n_i'(\vec x)$ are
polynomials of lower total degree than those of 
Eqn.~\eqref{eq:ansatz}.

We believe that a deeper understanding of these
two improvements might be fruitfully exploited
in the future to tackle more complex computations. 
In particular, it would be interesting to better
understand how the branch structure of pentagon functions
is related to the poles of the coefficients, and if this
suggests a superior basis of pentagon functions with
simpler coefficients.

\section{Implementation}

Let us now discuss how we applied the techniques of the previous
section to perform the analytic reconstruction.
We first compute numerical values for the
coefficients in a decomposition of one- and two-loop amplitudes
in terms of master integrals, using our C++ implementation of
numerical unitarity on a finite field 
\cite{Abreu:2017hqn,Abreu:2018jgq}. We then introduce
expressions for the master integrals in terms of pentagon
functions and compute the remainders in 
Eqn.~\eqref{eq:remainderHV} or \eqref{eq:remainderMHV}.
After changing the basis of pentagon functions and multiplying 
by the predetermined denominator structures, we arrive at an
evaluation of the numerator polynomials $n_i'(\vec x)$. 
Table~\ref{tab:ranks} summarizes the impact of the improvements,
the difficulty in the reconstruction of each amplitude,
and the number of different
letters appearing in the denominators of each helicity.
Given the simplicity of $\mathcal{R}_{\pm++++}$, we did not
implement any basis change for these helicities. We
observe a drastic increase in complexity from top to bottom,
as expected from one-loop amplitudes \cite{Bern:1993mq}.
Comparing the second and third columns we find that the
coefficients in Eqn.~\eqref{eq:dec_as_masters} are indeed 
simpler than those in Eqn.~\eqref{eq:ampAsPent}. 
We also show that not all letters contribute to the denominator
in Eqn.~\eqref{eq:ansatz}, even for the most complicated
helicity.

The reconstruction of the multivariate polynomials is done with
an in-house implementation of the algorithm presented in 
Ref.~\cite{Peraro:2016wsq}. 
For a detailed description we refer to the original article.
The algorithm \cite{Peraro:2016wsq} correlates the sampled 
phase space to the function it reconstructs, learning about
simplifications as it progresses. For this reason, it is not trivial
to parallelize. In our implementation this is addressed by
anticipating a suitable superset of the points for a set of
functions. This implementation is well adapted to evaluations on
modern computer clusters.

{\small
\begin{table}[]
\begin{center}
\begin{tabular}{|c|c|c|c||c|} \hline
    helicity  & $\tilde c_{k,i}(t)$ &   $r_i(t)$ &
    $n'_i(t)$ & $ w_j\textrm{'s}\,\,\textrm{in denominator}$\\
    \hline
    $+++++$   & $t^{34}/t^{28}$ &  $t^{10}/t^{4}$ & $t^{10}$
    &3   
    \\[.5mm]\hline
    $-++++$   & $t^{50}/t^{42}$ &  $t^{35}/t^{28}$  & $t^{35}$
    &14 
    \\[.5mm]\hline
    $--+++$  & $t^{70}/t^{65}$ &  $t^{50}/t^{45}$  & $t^{40}$
    &17
    \\[.5mm]\hline
    $-+-++$  & $t^{84}/t^{82}$ &  $t^{68}/t^{66}$  & $t^{53}$
    &20
    \\[.5mm]\hline
\end{tabular}
\end{center}
\caption{
Each $t^n/t^d$  denotes the total degree of
numerator ($n$) and denominator ($d$) of the most complex
coefficient for each helicity amplitude in the
decomposition of Eqn.~\eqref{eq:ampAsPent} (second column) or 
Eqn.~\eqref{eq:dec_as_masters} (third column). The fourth
column lists the highest polynomial we reconstruct. The final
column lists the number of letters $w_j(\vec x)$ that
contribute in the denominator of Eqn.~\eqref{eq:ansatz}.}
\label{tab:ranks}
\end{table}
}

In order to demonstrate the efficiency of our approach we now 
summarize the computational requirements for the numerical
reconstruction of the presented amplitudes.
The
dense nature of the algorithm implies that the complexity grows
as the number of terms in a polynomial of total degree 
$R$ in $n$ variables, ${R + n \choose n}$. 
The functions we reconstruct are dimensionless and thus only
depend on the four variables $x_0$, $x_1$, $x_2$ and $x_3$, see
Eqn.~\eqref{eq:twistorVars}.
For example, for a generic polynomial with $R=53$, we would
need around 400,000 evaluations.
In practice, we observe that the polynomials we
reconstruct are not completely generic and we require fewer
evaluations.
For the most complicated remainder, $\mathcal{R}_{-+-++}$, we 
require 237,098 phase-space points for the reconstruction of 
all the functions $ n'_i(x)$.
For $\mathcal{R}_{--+++}$, we require 85,979 phase-space points.
The average time for the numerical computation of the
remainder of an MHV amplitude in a finite-field is 4 minutes per
phase-space point.
The reconstruction of the remainders 
$\mathcal{R}_{\pm++++}$ is simpler.
We perform the computation on a finite field of cardinality
$\mathcal{O}(2^{31})$. 
For the MHV amplitudes we also evaluate on
a second finite
field and apply the Chinese remainder theorem. With this, 
we are able to rationally reconstruct the result.

\section{Results}
We have validated the approach described above
by computing the one-loop amplitudes to all orders in 
$\epsilon$ (extending the results of \cite{Bern:1993mq}).
We then used it to obtain the analytic
form of all planar five-gluon two-loop amplitudes in the
leading-color approximation, in the 't~Hooft-Veltman 
regularization scheme. Our results are included in a set of
ancillary files~\footnote{The files can be downloaded from
\href{https://www.qft.physik.uni-freiburg.de/ScatteringAmplitudes}{\texttt{five-gluon amplitudes}}.
}
which we now describe.
The files contain analytic expressions for the remainders of 
four helicity configurations---$\mathcal{A}_{\pm++++}$,
see Eqn.~\eqref{eq:remainderHV}, and 
$\mathcal{A}_{-\mp\pm++}$,
see Eqn.~\eqref{eq:remainderMHV}---from which any other helicity
configuration can be obtained by symmetry or permutation of the
momenta. These remainders are
written as a linear combination of pentagon functions,
similar to Eqn.~\eqref{eq:dec_as_masters} but with the adjusted
bases. This allows
to use the library provided in \cite{Gehrmann:2018yef} to
evaluate the remainders. We also include analytic expressions 
for the one-loop amplitudes, which are required to obtain the
two-loop amplitudes from their remainder. These are written as a
linear combination of master integrals in the one-loop
basis of Ref.~\cite{Giele:2008ve} and valid to all orders in
$\epsilon$. We provide expressions for the expansion of the
one-loop master integrals through order $\epsilon^2$, written in
terms of
pentagon functions, so that they can easily be combined with the
expressions for the remainders. Finally, we include a script 
that assembles all components to evaluate a two-loop amplitude.

The results we present are valid in the Euclidean region.
Since they are written in terms of pentagon 
functions~\cite{Gehrmann:2018yef} there is
minimal work to extend the results to all kinematic regions.
We checked our expressions against recently
computed two-loop amplitudes 
\cite{Badger:2013gxa,Badger:2017jhb,Gehrmann:2015bfy,
Badger:2018enw,Abreu:2017hqn,Abreu:2018jgq,Dunbar:2016aux}.
In the ancillary files we include numerical values for the
pentagon functions at the phase-space point of 
Ref.~\cite{Abreu:2018jgq} and a script that
numerically evaluates the amplitudes at that point.

\section{Conclusions}

We have presented the analytic form of a complete set of
planar two-loop five-gluon amplitudes in the leading-color
approximation. 
The amplitudes were obtained by reconstructing
the analytic expressions from numerical samples on finite
fields, computed in the framework of two-loop numerical 
unitarity. This allows to make the most of the resilience of
numerical calculations to handle intermediate steps and to only
target the analytic form of final expressions, which is
constrained by various physical properties. By focusing on
finite remainders, we reduced the complexity of the objects to
reconstruct. Efficiency of the reconstruction was further
enhanced by an a priori determination of the denominators and
changes of basis that considerably reduce the total degrees 
of the numerators of the coefficients. Through this process,
we reduced the calculation of the coefficients to the
reconstruction of a multi-variate polynomial of relatively low
total degree, rendering the most complex MHV amplitudes easily
reconstructible on a modern computer cluster. 

Our results include the first computation of the analytic
form of the five-gluon MHV amplitudes at two loops. They are an
important contribution to the
NNLO QCD corrections to three-jet 
production at hadron colliders. While in principle the numerical
evaluation of amplitudes is sufficient, 
the efficiency requirements for phase-space
integration over the final states are high due to helicity and
color sums. The  analytic expressions that
we provide will help to control both the evaluation times and 
the numerical stability of future phenomenological studies.

We expect our computational approach to greatly contribute
to formal developments in the study of scattering
amplitudes in quantum field theory, and to the new era of
precision QCD in high-energy physics.

\section{acknowledgments}

We thank V.~Sotnikov and M.~Zeng for many useful discussions.
We thank C.~Duhr for the use of his \Mathematica{} package
\texttt{PolyLogTools}.
The work of S.A.~is supported by the Fonds de la
Recherche Scientifique--FNRS, Belgium.
The work of J.D. and F.F.C. is supported by the Alexander 
von Humboldt Foundation,
in the framework of the Sofja Kovalevskaja Award 2014, 
endowed by the German Federal Ministry of Education and 
Research.
The work of B.P.~is supported by the French Agence Nationale
pour la Recherche, under grant ANR--17--CE31--0001--01.
The authors acknowledge support by the state of
Baden-W\"urttemberg through bwHPC.
S.A.~and B.P.~thank the Galileo Galilei Institute for
Theoretical Physics for the hospitality and the INFN for 
partial support during the final stages of this work.

\bibliography{main.bib}
\end{document}